\documentclass[aps,
longbibliography,notitlepage,reprint,superscriptaddress]{revtex4-2}
\usepackage[colorlinks,linkcolor=blue,citecolor=blue,urlcolor=red]{hyperref}
\usepackage{soul}
\usepackage{changes}
\usepackage{float}
\usepackage{CJK}
\usepackage{graphicx}
\usepackage{amsmath,amssymb,amsfonts}
\usepackage{siunitx}
\usepackage{soul}
\usepackage{chngcntr}
\usepackage{chemformula}
\usepackage{MnSymbol}
\usepackage{dsfont}	
\usepackage{appendix}
\usepackage{lineno}

\newcommand{\HU}{John A. Paulson School of Engineering and Applied Sciences, Harvard University, Cambridge, MA 02138, USA}
\newcommand{\QSE}{Quantum Science and Engineering, Harvard University, Cambridge, MA 02138, USA}

\begin{document}

\author{Yunxiang Song}
\email{ysong1@g.harvard.edu}
\affiliation{\HU}
\affiliation{\QSE}
\author{Pawan Ratra}
\affiliation{\HU}
\author{Danxian Liu}
\affiliation{\HU}
\author{Jiayu Yang}
\affiliation{\HU}
\author{Zhongshu Liu}
\affiliation{\HU}
\author{Urban Senica}
\affiliation{\HU}
\author{Salma Mohideen}
\affiliation{\HU}
\author{Mingjie Zhang}
\affiliation{\HU}
\author{Xudong Li}
\affiliation{\HU}
\author{Donald Witt}
\affiliation{\HU}
\author{Joshua Mornhinweg}
\affiliation{\HU}
\author{Norman Lippok}
\affiliation{\HU}
\author{Eric Mazur}
\affiliation{\HU}
\author{Federico Capasso}
\email{capasso@seas.harvard.edu}
\affiliation{\HU}
\author{Marko Lončar}
\email{loncar@g.harvard.edu}
\affiliation{\HU}

\title{Ultrafast programmable Bragg reflection in photonic integrated circuits}

\begin{abstract}
Distributed Bragg reflectors (DBRs) are foundational building blocks of classical and quantum photonic technologies. However, their optical responses are typically fixed upon fabrication, limiting circuit robustness, reconfigurability, and functionality in applications from high-speed communications to quantum computing. Here, we demonstrate photonic chip-based programmable DBRs at telecommunications wavelengths, which are formed by electro-optically inducing refractive index contrast between periodic ferroelectric domains in thin-film lithium niobate waveguides. We achieve voltage-controlled Bragg reflection from zero to near-unity, and gigahertz-speed reflectivity modulation. Our results bring DBRs into the ultrafast programmable regime, opening new opportunities in topological photonics, cavity quantum electrodynamics, integrated lasers, and optical interconnects. The interplay between nanoscale ferroelectric domain engineering and strong electro-optic nonlinearity establishes a new design strategy for nanophotonic devices, otherwise inaccessible in bulk media.
\end{abstract}

\maketitle

\begin{figure*}[t]
  	\includegraphics[width = 1.0\textwidth, page = 1]{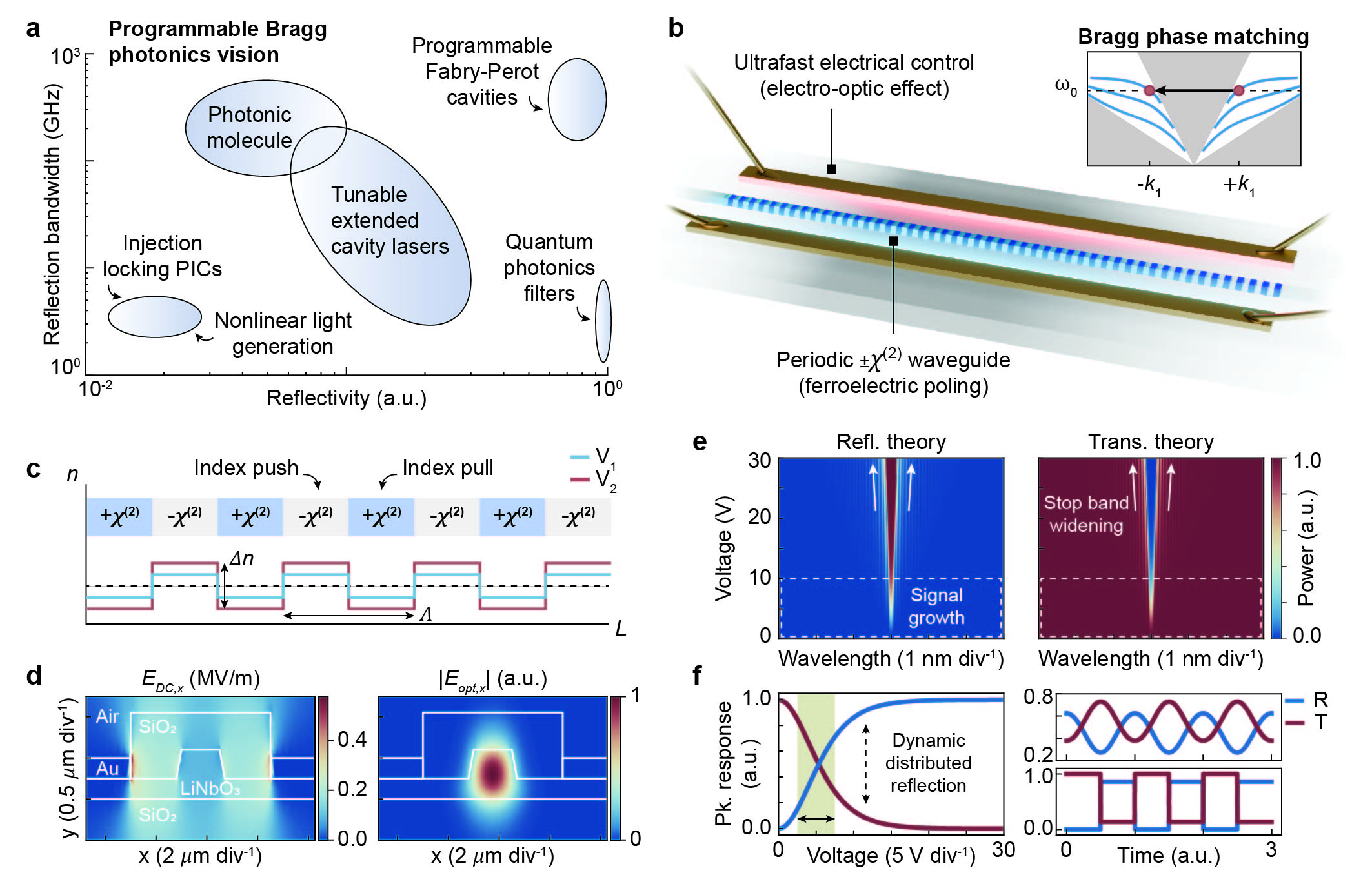}	\caption{\textbf{Ultrafast programmable Bragg photonics vision and device concept.} \textbf{(a)}, Envisioned application space of ultrafast programmable Bragg photonics across various reflectivity and reflection bandwidth combinations. Concrete examples of applications that would be enhanced by tunable distributed Bragg reflection are listed. PICs: photonic integrated circuits. 
    \textbf{(b)}, Schematic illustration of the electro-optic (EO) nonlinear ferroelectric grating, formed by EO tuning of a waveguide with periodically-inverted second-order susceptibility ($\chi^{(2)}$). Alternating ferroelectric domains of the waveguide are produced by periodic poling and shown in white and blue. The waveguide does not have etched-in structural perturbations. Top right inset shows a schematic of the dispersion curve for the guided modes of the waveguide, and the black arrow indicates the phase mismatch between counterpropagating guided modes. $\omega_0=2\pi c/\lambda_0$ is the angular frequency, where $c$ is the speed of light and $\lambda_0$ is the target wavelength of Bragg phase matching. $\pm k_1$ are wavevectors of the coupled modes evaluated at $\omega_0$. 
    \textbf{(c)}, Schematic illustration of the push-pull refractive index grating, under zero bias (black dashed line) and finite bias (blue and red lines). Larger index contrast ($\Delta n$) between unit cells in a grating period ($\Lambda$) is achieved by applying a larger voltage ($V_2>V_1$). 
    \textbf{(d)}, Left panel shows the simulated electric field distribution for an applied voltage of one volt, and assuming the same physical parameters considered in our devices. Right panel shows the fundamental transverse-electric-like optical mode profile. Au: gold, SiO$_2$: silicon dioxide, LiNbO$_3$: lithium niobate. The top SiO$_2$ cladding is 800 nm-thick, the Au electrode is 250 nm-thick, the LiNbO$_3$ waveguide consists of 350 nm ridge height and 250 nm slab height, and the waveguide’s sidewall angle is 60 degrees. 
    \textbf{(e)}, Simulated reflection and transmission spectra as a function of applied voltage $V$, assuming a voltage-dependent coupling constant $\kappa=\alpha\cdot V$ between counterpropagating modes, where $\alpha$ is the coupling constant per unit voltage. The reflected signal rapidly grows within the first 10 V, and the bandwidth broadens as the peak reflectivity asymptotically approaches unity for higher voltages. 
    \textbf{(f)}, Left panel shows the reflected (blue) and transmitted (red) power vs. applied voltage at the Bragg wavelength (Pk: peak), and right panels illustrate analog and digital modulation of the DBR response across the shaded region.}
	\label{fig:fig1}
\end{figure*}

\begin{figure}[!htbp]
    \includegraphics[width = 0.48\textwidth, page = 1]{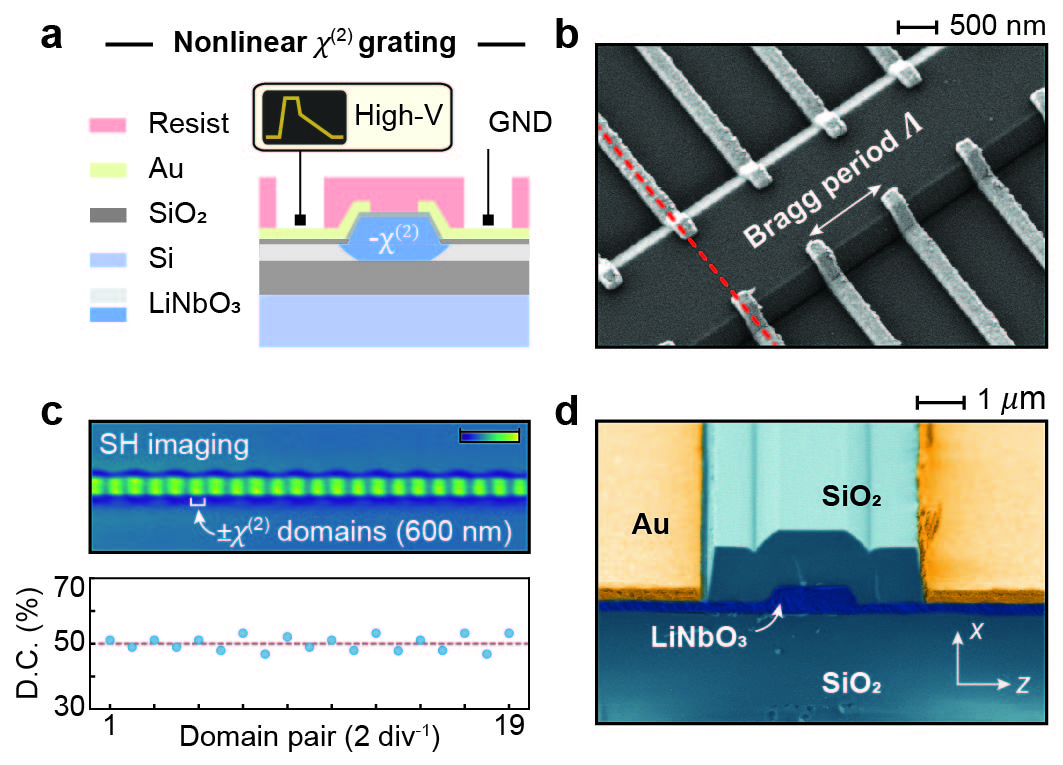}  
    \caption{\textbf{Realization of electro-optic nonlinear ferroelectric gratings on thin-film lithium niobate.}
    \textbf{(a)}, Schematic of the periodic poling process that forms the nonlinear $\pm\chi^{(2)}$ grating. The specific cross section shown corresponds to the red dashed line in panel (b). A high-voltage (High-V) pulse is delivered to the poling electrodes, resulting in ferroelectric switching of the crystal orientation in the corresponding region (labeled by $-\chi^{(2)}$ and colored in dark blue). Au: gold, SiO$_2$: silicon dioxide, Si: silicon, LiNbO$_3$: lithium niobate, GND: electrical ground. 
    \textbf{(b)}, Scanning electron microscope image of the poling electrodes (bright fingers), positioned along both sidewalls of the optical waveguide (dark rectangle). The poling period is the desired distributed Bragg reflector period, $\Lambda$. Cross section along the red dashed line is shown in panel (a). 
    \textbf{(c)}, Top panel shows the confocal second harmonic (SH) microscope image of a poled waveguide, where the color map represents the SH signal intensity. Periodic $\pm\chi^{(2)}$ domains in the waveguide are separated by vertical lines that mark the domain boundaries. The SH signal is suppressed at these boundaries because SH fields emitted from adjacent $\pm\chi^{(2)}$ domains are out of phase and destructively interfere. Comparable signal intensities (bright green) within both types of domains are consistent with near-complete domain inversion. Bottom panel shows the domain duty cycle (D.C.) extracted from the top panel image and plotted as blue dots, achieving roughly 50\% (red dashed line). 
    \textbf{(d)}, False-colored scanning electron microscope image of the final device cross section. The bottom right inset marks the axes of the lithium niobate crystal.}
  \label{fig:fig2}
\end{figure}

Modern integrated photonic systems, from high-speed optical interconnects \cite{sun2015single, rizzo2023massively} to photonic quantum computers and information processors \cite{psiquantum2025manufacturable, zhang2025classical}, are fundamentally built from the most basic photonic components such as waveguides, resonators, and gratings. These components are traditionally considered passive, since their optical properties are largely fixed upon fabrication. However, as integrated photonics continues to scale towards increasingly large and multifunctional circuits, static individual components become a major limitation.

Overcoming this limitation in future photonic systems will require active programmability down to the component level, enabling circuits that are robust to fabrication variations as well as reconfigurable in their function. Recent advances in programmable photonic circuits \cite{bogaerts2020programmable, wu2023lithography, yanagimoto2026programmable} include reconfigurable waveguides and resonators, featuring tunable optical path length and dispersion, as well as resonator size and coupling rate \cite{perez2017multipurpose,harris2018linear}. These programmable building blocks have expanded the capabilities of microwave photonics \cite{marpaung2019integrated, perez2024general}, optical computing \cite{shen2017deep, mcmahon2023physics}, topological photonics \cite{dai2024programmable, ma2026reconfigurable}, and photonic quantum information processing \cite{nielsen2025programmable, aharonovich2026programmable}. Despite these advances, comparable levels of programmability remain elusive to optical gratings, another ubiquitous component in photonic integrated circuits.

Optical gratings use spatially-periodic refractive index modulations to phase match and transfer power between the otherwise phase-mismatched optical fields propagating within them. Reflective gratings that couple counterpropagating fields, also known as distributed Bragg reflectors (DBRs), underpin many key devices in integrated photonics, such as filters \cite{wang2012narrow}, wavelength and mode multiplexers \cite{dai2016silicon}, Fabry-Perot resonators \cite{wildi2023dissipative, hwang2025high}, photonic molecules \cite{arbabi2011realization, lu2022high}, nonlinear optical cavities \cite{yu2021spontaneous, ulanov2024synthetic}, and continuous-wave and mode-locked lasers \cite{xiang2019ultra, qiu2026high}. The DBRs involved are often static, since the periodic index profiles are lithographically defined and etched into the underlying material. Actively reconfigurable DBRs, on the other hand, would dramatically improve device robustness and flexibility, while also unlocking new possibilities for the applications outlined above [Fig. 1a].

Realizing actively reconfigurable DBRs would require refractive index control at subwavelength spatial scales, yet conventional index-tuning mechanisms face fundamental spatial-resolution limits. Thermo-optic tuning is constrained by thermal diffusion, while electro-optic tuning is limited by electric-field fringing, restricting practical index modulation to length scales of tens of micrometers. As a result, existing DBR tuning has been restricted to global refractive index changes \cite{zhang2018fully, prencipe2021tunable, pohl2020100, siddharth2025ultrafast, xue2025pockels}. Although phase-change materials have been proposed for programming subwavelength refractive index profiles, their chip-scale integration is difficult and experimental demonstrations are limited \cite{nobile2023nonvolatile}. More recently, suspended opto-electro-mechanical devices have achieved megahertz-rate, subwavelength refractive index tuning in silicon, albeit operating in the mid-infrared and facing potential limitations for further wavelength scaling \cite{liu2025ultracompact}. To date, a concept for programmable DBRs in photonic integrated circuits, that simultaneously feature low fabrication complexity, full-dynamic-range and high-speed reconfigurability, as well as near-infrared compatibility, is missing.

In this work, we address this challenge by demonstrating electrically programmable distributed Bragg reflection at telecommunications wavelengths, in a planar thin-film lithium niobate (TFLN) waveguide. This functionality is achieved by combining the electro-optic (EO) effect with a nonlinear ferroelectric grating formed by periodic poling [Fig. 1b]. A core innovation of our approach is that the index modulation resolution is set by domain reversal, which, in thin films, is compatible with deeply subwavelength spatial profiles \cite{yang2024symmetric, sabatti2025nanodomain}. Additionally, the absence of any macroscopic structural perturbations, such as sidewall corrugations, allows the ferroelectric gratings to preserve the low-loss properties of TFLN, while index modulation based on electro-optics enables reflectivity switching exceeding gigahertz-speeds. Our results establish a foundation for ultrafast programmable Bragg photonics in a scalable optoelectronic platform \cite{hu2025integrated}, central to next-generation communications and computing technologies.

\begin{figure*}[!t]
    \includegraphics[width = 1.0\textwidth, page = 1]{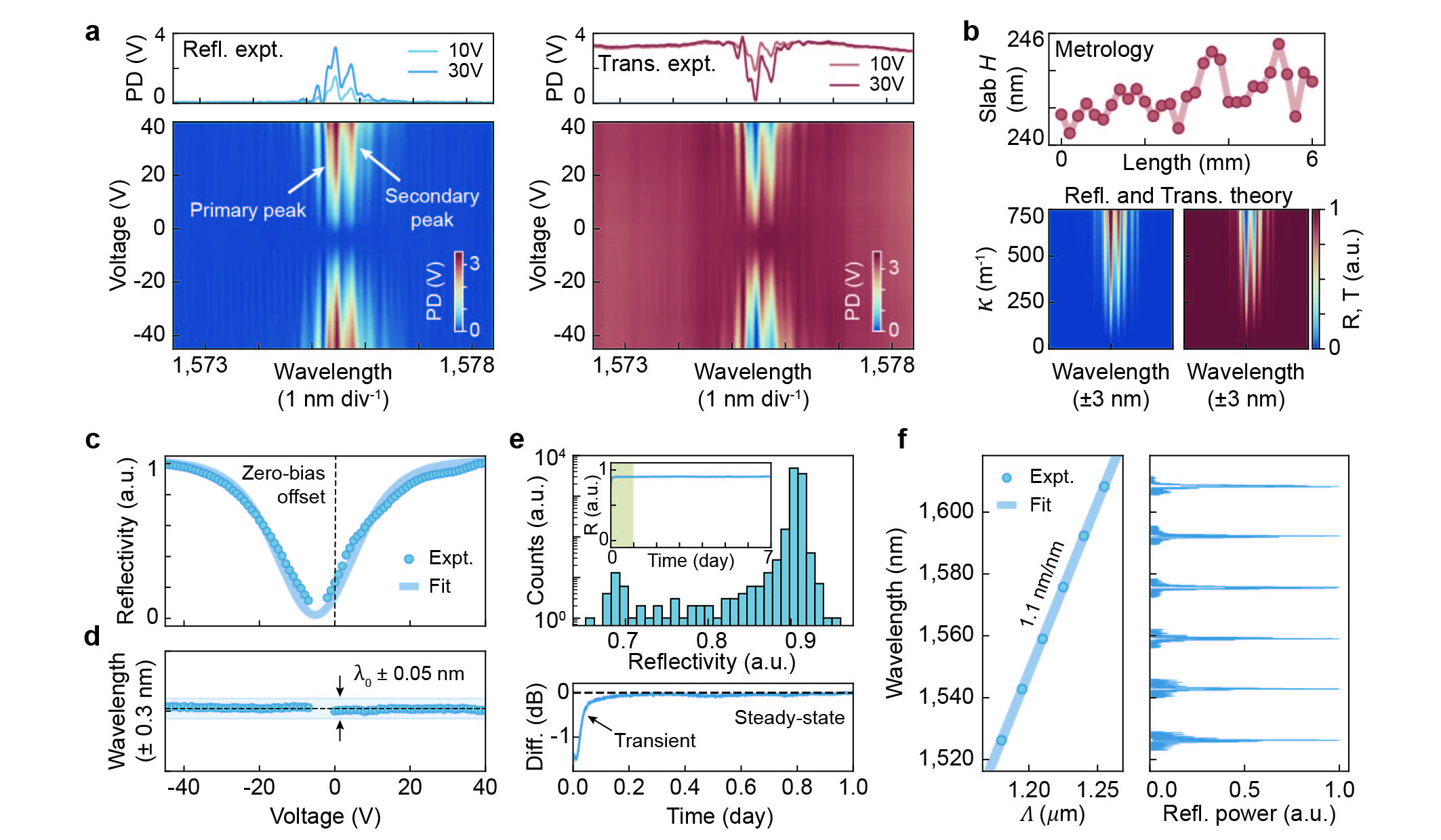}  
    \caption{\textbf{Electrically programmable distributed Bragg reflection.} \textbf{(a)}, Left (right) panel shows the reflection (transmission) spectrum of an $L=6$ mm-long programmable distributed Bragg reflector (DBR), under bias voltages from -45 V to 40 V in 1 V increments. A primary and secondary reflection peak are marked by the white arrows, and deviation from ideal transfer functions are due to film thickness variations along the DBR. Representative spectra under 10 and 30 V-bias are plotted in the top sub-panels. PD: photodetector. 
    \textbf{(b)}, Top panel shows the waveguide slab height, $H$, measured at 200 $\mu$m intervals along the device. Assuming that the film thickness follows the same spatial variation, the simulated reflection ($R$) and transmission ($T$) spectra are shown in the bottom panels as functions of the coupling constant $\kappa$. The simulations accurately reproduce both the double-peaked transfer function and the relative peak amplitudes. 
    \textbf{(c)}, Reflectivity vs. voltage (blue dots). 
    The experimental data is fitted to a tanh$^2\big(\alpha\cdot(V-V_o)\cdot L\big)$ model, where the coupling constant $\kappa$ is expressed as $\alpha\cdot(V-V_o)$, $\alpha$ is the coupling constant per unit voltage, $V$ is the bias voltage, and $V_o$ is the offset voltage. Assuming $L=6$ mm, we find the direct-current (DC) electro-optic (EO) efficiency is given by $\alpha\approx11.95$ (m$\cdot$V)$^{-1}$ and $V_o\approx-5.16 $V. The fitted $\alpha$ is lower than the simulated efficiency $\alpha=18.34$ (m$\cdot$V)$^{-1}$ for an ideal grating (adjusted for film thickness nonuniformity); other sources of discrepancy are discussed in further detail in the Supplementary Information. 
    \textbf{(d)}, Bias voltage vs. measured DBR center wavelength (blue dots), indicating stable center wavelength and a balanced push-pull index grating. Light blue area demarcates $\pm 0.05$ nm around the Bragg wavelength $\lambda_0$, which is the approximate 3-dB bandwidth of the primary reflection peak. The gaps in the data in (c) and (d) are conditions of near-zero reflectivity and thus the reflectivity and $\lambda_0$ cannot be accurately identified. 
    \textbf{(e)}, Top panel shows a histogram of reflectivities under constant 20 V-bias, monitored approximately once per minute over seven days, with counts plotted on a log scale. Inset shows the corresponding reflectivity vs. time and shows stable operation. The small number of low-reflectivity counts is due to the initial transient response that is expanded in the bottom panel, and this response eventually reaches steady-state (black dashed line). Diff: difference from steady-state value of $R\approx90$\%. 
    \textbf{(f)}, Left panel shows the DBR center wavelength vs. Bragg period $\Lambda$ for six devices. The experimental data (blue dots) are well described by a linear fit (blue line), with a slope of 1.1 nm$\cdot$nm$^{-1}$. Right panel shows representative reflection spectra of the devices under 50 V bias, from which the center wavelengths in the left panel were extracted.
    }
  \label{fig:fig3}
\end{figure*}

\section{Device concept and design}

Figure 1b illustrates our concept: coplanar electrodes supply a control electric field across a periodically-poled TFLN waveguide. Periodic poling of lithium niobate produces domains with alternating signs of second-order nonlinear susceptibility ($\chi^{(2)}$), historically developed for quasi-phase matching in nonlinear optics \cite{armstrong1962interactions}. Here we exploit the corresponding periodic sign reversal of the EO coefficient, which is proportional to $\chi^{(2)}$, to form a programmable refractive index grating under an applied electric field.

The poling period, $\Lambda=\lambda_0/2n_{\text{eff}}(\lambda_0)$, is chosen to Bragg-phase-match counterpropagating waves at the center wavelength $\lambda_0$, where $n_\text{eff}$ is the effective index of these waves. At zero bias field, the device behaves as an ordinary low-loss waveguide, since the linear refractive index is uniform [Fig. 1c, black dashed line]. When a bias field is created through an applied voltage, the alternating EO response between domains produces a spatially periodic refractive index change ($+\chi^{(2)}$ domains experience decreasing index and vice versa), transforming the waveguide into a DBR. Increasing the bias field increases the index contrast $\Delta n$ [Fig. 1c, compare $V_1$ and $V_2$ curves], which strengthens the coupling constant $\kappa$ between the counterpropagating waves. Specifically, $\Delta n$ (and hence $\kappa$) is proportional to the overlap integral between the localized electric and optical fields [Fig. 1d]; see Supplementary Information for details.

To illustrate the expected tuning efficiency, we perform numerical simulations using realistic design parameters considered in the experiments described below [Fig. 1d, caption]. Defining a linear relationship via the EO effect, $\kappa=\alpha\cdot V$, where $\alpha$ is the coupling constant per unit voltage and $V$ is the applied voltage, we obtain $\alpha=82.62$ (m$\cdot$V)$^{-1}$ for an ideal first-order DBR. In our implementation, the programmable DBR is poled at third-order ($\Lambda=3\lambda_0/2n_\text{eff} (\lambda_0)\approx1.2$ $\mu$m for $\lambda_0\approx1,550$ nm), which reduces $\alpha$ by a factor of three. Thus, we use $\alpha=27.54$ (m$\cdot$V)$^{-1}$ to simulate reflection and transmission spectra, assuming an $L=6$ mm-long device [Fig. 1e]. Despite the use of a higher-order DBR, the reflected signal increases rapidly within the first 10 V, while the stopband gradually broadens as the applied voltage is further increased. This behavior is consistent with conventional DBRs as their index contrast is smoothly increased. Plotting the reflected and transmitted power transfer functions at $\lambda_0$ as a function of voltage [Fig. 1f], we identify a region where the DBR response is strongly voltage-sensitive, and may be used for both analog (arbitrary waveform) and digital (on-off switching) reflection modulation, depending on the application.

\section{Electro-optic nonlinear ferroelectric grating on thin-film lithium niobate}

The EO nonlinear ferroelectric gratings are fabricated on congruent X-cut TFLN-on-insulator wafers. After etching the optical waveguides, we formed nonlinear $\pm\chi^{(2)}$ gratings by periodic poling. Poling electrodes are placed along the waveguide sidewalls, and a single high-voltage pulse ($\approx 190$ V peak) is applied to induce alternating ferroelectric domains at the DBR period [Fig. 2a, 2b]. After poling, these electrodes were removed, and the poling quality was evaluated using confocal second-harmonic microscopy: in Fig. 2c, the $\pm\chi^{(2)}$ domains along the waveguide are clearly separated by vertical dark lines (domain walls), and their comparable brightness indicates a large depth of domain inversion. The domain duty cycle of roughly 50\% is further extracted from the microscope image. We note that the poling electrode geometry and high-voltage pulse shape were co-optimized to achieve near-complete domain inversion and a uniform duty cycle at subwavelength scales, thereby maximizing the tuning efficiency of the ultrafast programmable DBR. The remaining fabrication steps followed standard TFLN EO modulator workflows for top cladding and metallization. Figure 2d shows the cross section of a completed device, in which electrode pads flank both sides of the $\pm\chi^{(2)}$ waveguide. Full fabrication details and device parameters are provided in the Supplementary Information.

\section{Static field-programmable distributed Bragg reflection}
After verifying that the fabricated devices matched the proposed concept, we first characterized their programmability under static electric fields. For an $L=6$ mm-long device with third-order Bragg period $\Lambda=1.225$ $\mu$m (0.009 mm$^2$ footprint), we sweep the bias voltage from -45 V to 40 V in 1 V increments. The reflection and transmission spectra (collected by grating coupling) are shown in left and right panels of Fig. 3a, respectively. With moderate voltages of up to 30 V, the reflectivity at the Bragg wavelength can be programmed from near-zero to near-unity. In addition to observing clear reflectivity growth and bandwidth broadening with increased voltage, two additional features emerged: a less efficient, secondary reflection peak red-shifted from the primary one, and a small, passive reflectivity under zero bias voltage.

We attribute these imperfections in the transfer function to a nonuniform initial film thickness, which changes the Bragg condition along the device. This hypothesis is confirmed by film thickness metrology, where the waveguide thickness is mapped in 200 $\mu$m steps [Fig. 3b, top panel]. Accounting for this nonuniformity, our simulations now reproduce the double-peaked transfer function [Fig. 3b, bottom panel]. This sensitivity to film thickness ($\pm3$ nm in this device) is not unique to our grating design; rather, it is a property of all DBRs. This thickness variation also contributes to the discrepancy between simulated and measured tuning efficiencies by effectively reducing the grating length. Accounting for this reduction lowers the theoretical $\alpha$ from $27.54$ (m$\cdot$V)$^{-1}$ to $18.34$ (m$\cdot$V)$^{-1}$, bringing it closer to the measured response in Fig. 3a compared to the idealized simulations in Fig. 1e. Because these imperfections can be accurately predicted through metrology, adaptive grating designs \cite{chen2024adapted} will be effective if perfect transfer functions are necessary. We also note that while the double-peaked transfer functions may intuitively suggest two dominant periodic components in the nonuniformity, they instead arise from the effect of cascaded nonidentical DBRs coupled together (see Supplementary Information for details).

Next, we discuss the phenomenon of non-zero passive reflectivity. Comparing DBR responses of devices poled with different high-voltage pulse shapes (which varies the domain duty cycle), we found that the zero-bias background signal correlates with deviations from a 50\% poling duty cycle (see Supplementary Information for extended analysis). Although poling reverses only the ferroelectric polarization and is not expected to directly alter the linear refractive index, the periodic domain walls along waveguide edges may be accompanied by a weak internal electric field induced by space-charge accumulation \cite{nataf2020domain}. Nevertheless, these offsets are not fundamental and can always be compensated using additional bias voltage: in the present device, the peak reflected power follows the expected $\text{tanh}^2\big(\alpha\cdot(V-V_o)\cdot L\big)$ scaling law for both positive and negative applied voltages, with an offset voltage of $V_o\approx-5.16$ V [Fig. 3c]. The center wavelength shows no appreciable shift with bias voltage, indicating a nearly balanced push-pull index change across the $\pm\chi^{(2)}$ domains [Fig. 3d]. We note that the fitted $\alpha\approx 11.95$ (m$\cdot$V)$^{-1}$ is still lower than the nonuniformity-adjusted (effective) theoretical $\alpha\approx 18.34$ (m$\cdot$V)$^{-1}$. This discrepancy is likely due to residual deviations from a 50\% poling duty cycle and incomplete domain inversion.

To evaluate the device’s bias stability, we performed a seven-day measurement under a constant 20 V bias. Transmission and reflection spectra are collected approximately every minute, and the former is used to track the reflectivity over time. We observed a brief transient period after 20 V was applied, after which the DBR response reached a steady state [Fig. 3e, bottom panel]. This response remained strong throughout the measurement duration, showing excellent long-term stability. For clarity, we plot all sampled reflectivities in a histogram (logarithmic vertical scale), displaying a dominant peak around the steady-state response of $R\approx90$\% [Fig. 3e, top panel].

Finally, we examine additional $L=6$ mm-long devices with different poling periods. Increasing the period from $\Lambda=1.180$ $\mu$m to $\Lambda=1.255$ $\mu$m shifts the DBR center wavelength from $\lambda_0\approx1,526$ nm to $\lambda_0\approx1,608$ nm [Fig. 3f; all devices biased at 50 V]. The strong linear dependence of $\lambda_0$ on $\Lambda$ is consistent with simulations and demonstrates that programmable DBRs with tailored spectral responses can be realized through an appropriate set of spatial frequencies.

\begin{figure*}[!htbp]
    \includegraphics[width = 1.0\textwidth, page = 1]{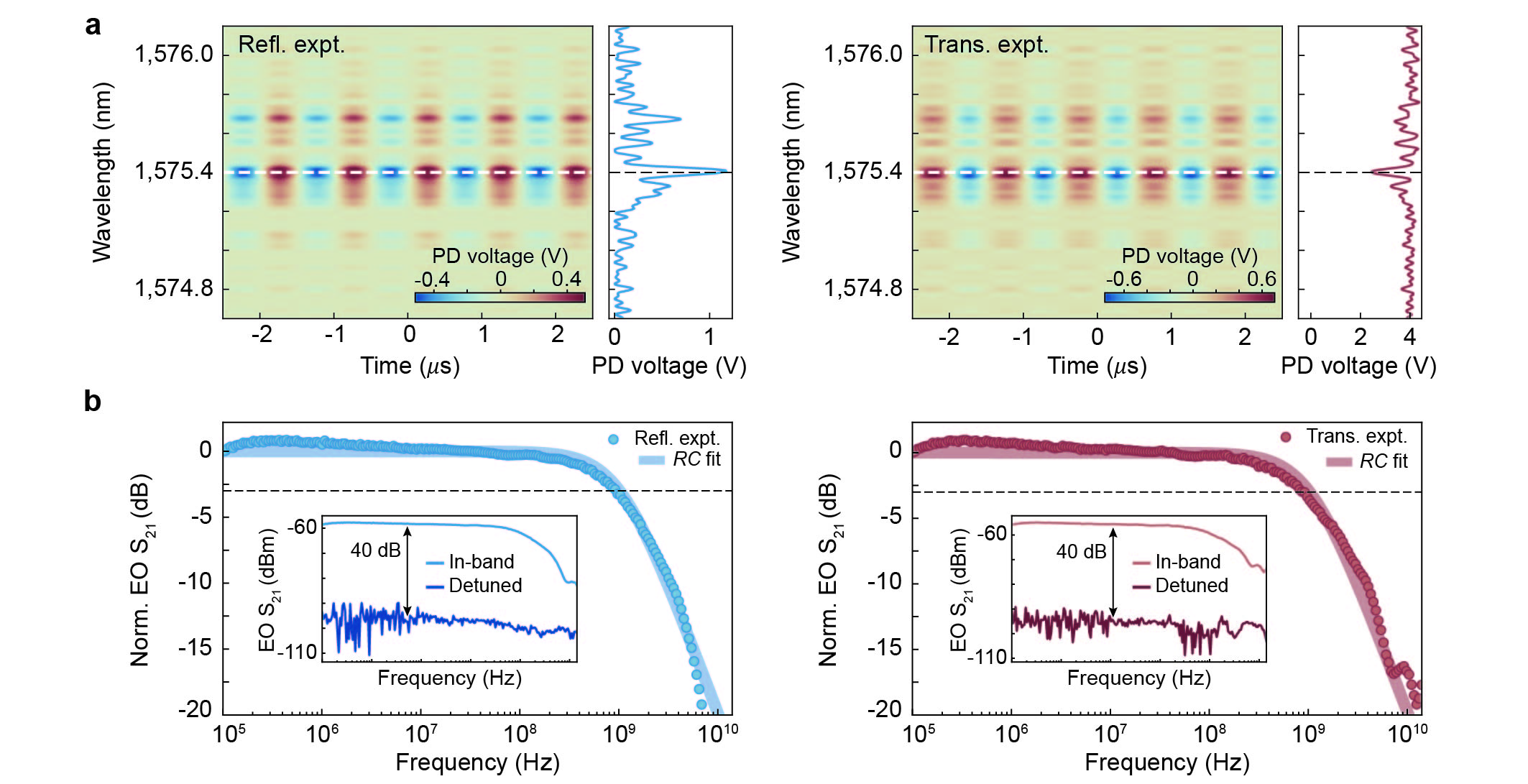}  
    \caption{\textbf{High-speed electrical modulation of distributed Bragg reflection. } 
    \textbf{(a)}, AC-coupled modulation response vs. laser wavelength maps in reflection (left panel) and transmission (right panel). The modulation signal is a 1 MHz sine wave with 5 V amplitude and 5 V bias. In each panel, the right subpanels show the raw reflection and transmission spectrum. Additional small-amplitude oscillations in the reflection spectra, outside the distributed Bragg reflector (DBR) features, arise from the weak Fabry-Perot background formed by the grating couplers, which were filtered out in Fig. 3a for clarity. The modulation response is strongest when the laser wavelength is aligned with the DBR features, including the primary and secondary reflection peaks identified in Fig. 3a. In contrast, there is negligible response elsewhere, confirming the modulation origin is from dynamic distributed Bragg reflection. Dashed lines label 1,575.4 nm and serve as guides to the eye. 
    \textbf{(b)}, Normalized frequency response (electro-optic $S_{21}$) in reflection (left panel) and transmission (right panel) for a fixed laser wavelength of 1,575.42 nm, as measured by a vector network analyzer. The frequency sweep spans 201 values between 100 kHz and 14 GHz, equally-spaced in log scale. The experimental data (dots) are fitted to $RC$-limited response curves, with fitted 3-dB bandwidths of 1.024 GHz and 1.035 GHz in reflection and transmission, respectively. Black dashed lines label -3 dB. Insets show that when the laser wavelength is detuned away from the primary reflection peak by 1 nm, the absolute frequency response drops to the vector network analyzer noise floor, again confirming the physical origin of the modulation.}
  \label{fig:fig5}
\end{figure*}

\section{Ultrafast control of distributed Bragg reflection}
The experiments above clearly demonstrate the static field-programmability of our devices.  Since the gratings are fundamentally based on electro-optics, we can further expect ultrafast electrical modulation across their characteristic response maps [Fig. 3a]. To this end, we now examine our device’s performance under high-frequency microwave fields.

First, we apply a 1 MHz sinusoidal modulation (5 V amplitude and 5 V bias) to the device considered in Fig. 3a. As the modulation is applied, we sweep the laser wavelength across the primary reflection peak ($\lambda_0\approx 1,575.40$ nm), from $1,574.65$ nm to $1,576.15$ nm in $0.001$ nm steps. The reflected and transmitted laser powers are strongly modulated only when the laser wavelength is aligned with the DBR features [Fig. 4a]. This includes modulation of both the primary and secondary reflection peaks, although the modulation swing of the secondary peak is weaker due to its comparatively shorter interaction length.

Next, we measure the device’s high-frequency response at a fixed laser wavelength within the primary reflection peak ($1,575.42$ nm) using a vector network analyzer. For both reflected and transmitted signals, the frequency response (EO $S_{21}$) remains relatively flat starting from 100 kHz with 3-dB roll-offs near 1 GHz, as fitted by $RC$-limited response curves [Fig. 4b]. Consistent with measurements in Fig. 4a, the grating’s EO response drops to the vector network analyzer noise floor when the laser is optically detuned to $1,576.40$ nm (arbitrarily chosen), confirming that the modulation arises from dynamic distributed Bragg reflection [Fig. 4b, insets].

\section{Discussion}

We presented EO nonlinear ferroelectric gratings that enable efficient tuning of refractive index at small spatial scales. Surpassing conventional DBR tuning, which only shifts the center wavelength, we experimentally demonstrated a new regime of programmability in photonic-integrated DBRs: electrical activation of distributed Bragg reflection and gigahertz-speed reflectivity modulation. We envision that our results will lay the foundation for next-generation programmable photonics based on reconfigurable and fast-switchable grating structures, with immediate applications in established grating-based technologies and beyond [Fig. 1a].

At telecommunications and other near-infrared wavelengths, our devices may unlock new classes of integrated lasers by harnessing on-chip rare-earth \cite{chen2021efficient, liu2022photonic, singh2025watt} and semiconductor gain media \cite{xiang2019ultra, guo2023ultrafast}, featuring in situ output power optimization and Q-switching. Expanded capabilities to generate high-power continuous-wave and pulsed lasers in chip-scale formats should enable novel sources for optical transceivers and frequency combs for precision ranging and nonlinear science. Additionally, temporal EO modulation combined with the periodic $\chi^{(2)}$ spatial profile of our gratings could realize space–time modulation physics in a nanophotonic platform, accessing exotic effects such as magnetic-free nonreciprocity \cite{lira2012electrically, sounas2017non}, and optical field amplification and compression \cite{horsley2024traveling}. Although the gratings are currently third-order poled, their spatial resolution is not fundamentally limited: periodic poling in ferroelectric thin-films can achieve periods as small as $\approx200$ nm \cite{yang2024symmetric, sabatti2025nanodomain, franken2026milliwatt}. This capability should extend our approach to first-order poled devices at telecommunications wavelengths with enhanced efficiency and bandwidth, as well as shorter near-infrared and even visible wavelength devices compatible with heterogeneously integrated color centers \cite{riedel2023efficient, ding2024high} and quantum dots \cite{davanco2017heterogeneous, chen2026chip}. In such contexts, programmable DBRs could enable photonic crystal cavities with dynamically modulated light-matter interaction strength and photon storage and routing, opening new opportunities for on-chip cavity quantum electrodynamics and quantum information science. Finally, recent successes in the heterogeneous integration of thin-film ferroelectrics with other materials \cite{churaev2023heterogeneously, niels2026high, brodnik2026monolithic} may broaden the scope of our results beyond TFLN, unlocking ultrafast programmable Bragg reflection in a variety of photonics platforms.

\section*{Acknowledgments}
We thank Jinsheng Lu, Andrea Cordaro, Aaron Day, Kees Franken, Evelyn Hu, Shengyuan Lu, Hana Warner, and Matthew Yeh for discussions. 

\section*{Contributions}
Y.S. conceived the idea for the project. Y.S. designed the devices and performed the simulations, with P.R. assisting. Y.S. developed the fabrication process and fabricated the devices, with D.L., J.Y., Z.L., and D.W. assisting. J.Y. provided input on periodic poling. Y.S. performed the experiments, with D.L., P.R., Z.L., U.S., S.M., and N.L. assisting. M.Z., X.L., and J.M. provided additional help. Y.S. analyzed the data and wrote the manuscript, with substantial contributions from F.C. and M.L.; all authors commented on the manuscript. All authors discussed the results. E.M., F.C., and M.L. supervised the project.

\section*{Funding}
We acknowledge support from Air Force Office of Scientific Research (FA955025CB011, FA9550-23-1-0699), Defense Advanced Research Projects Agency (DARPA HR0011-24-2-0360), National Institute of Health (5R21EY037019-02), Swiss National Science Foundation (P500PT 214483 1), Naval Sea Systems Command (N0002425CT146), National Science Foundation (EEC-1941583), and Amazon Web Services (A60290).

\section*{Competing interests}
M.L. is involved in developing lithium niobate technologies at HyperLight Corporation. The authors declare no other competing interests.

\section*{Data availability}
All data needed to evaluate the conclusions in the paper are present in the paper and/or the Supplementary Information.




\bibliographystyle{apsrev4-2}
\bibliography{refs}

\end{document}